\documentstyle[psfig,12pt,twocolumn]{article}

\begin{document}

\setlength{\textwidth}{480pt}
\setlength{\textheight}{630pt}
\setlength{\topmargin}{-0.375in}
\setlength{\oddsidemargin}{-0.0833in}
\setlength{\evensidemargin}{-0.0833in}
\setlength{\parindent}{0in}
\setlength{\parskip}{0pt}

\title{Fundamental Disagreement of Wave Mechanics with Relativity}

\author{
{\large Ezzat G. Bakhoum}\\
\\
{\normalsize New Jersey Institute of Technology}\\
{\normalsize P.O. Box 305, Marlton, NJ. 08053 USA}\\
{\normalsize Email: bakhoum@modernphysics.org}\\
\\
{\normalsize (This work has been copyrighted with the Library of Congress)}\\
{\normalsize Copyright \copyright 2001-2002 by Ezzat G. Bakhoum}\\
{\normalsize Journal Ref: Physics Essays, Vol.15, No.1}
}

\date{\normalsize Posted: June 17, 2002 \hspace*{.25in} Updated: December 14, 2002}

\maketitle

{\large\bf Abstract:}\\
\\
A number of well-known difficulties in physics resulted from merging the theory of relativity with the Compton-de Broglie wave mechanics. Two such problems were  the failure of Dirac's relativistic wave equation to predict the correct velocity of the electron, and the fact that the measured yield from nuclear fission was found to be substantially less than the theoretical yield. It is shown that the origin of these and other problems stem from the inconsistency of the relativistic mass-energy equivalence principle with the fundamental assumptions of wave mechanics. An alternative view of the concept of mass-energy equivalence that results in a very good agreement between theory and experiment is demonstrated. The conclusions of this paper will be quite important for ongoing research, such as the current problem of the neutrino's mass.

\pagebreak

{\large\bf 1. Introduction:}\\
\\
The success of the modern theory of matter is certainly due in no small part to its dependence on the new wave mechanics of Compton, de Broglie and Schr\"{o}dinger. As it is firmly established, the modern theory of matter  began with the independent efforts of Bohr and Sommerfeld and was subsequently refined  with the introduction of the non-relativistic Compton-de Broglie mechanics. Later, however, efforts to merge that new theory with the theory of relativity resulted in some successes and some serious difficulties, some of which remain largely unsolved today. The first difficulty was the mystery of the lost energy in the beta-decay process, for which W.Pauli postulated the existence of the neutrino. While the existence of that particle was verified experimentally, any calculations involving the neutrino's energy or mass remain inconclusive\cite{Schwarzschild},\cite{Bergkvist}. Another difficulty was encountered by Dirac\cite{Dirac} in his attempt to obtain a relativistic wave equation to describe the motion of a free electron. Dirac obtained a Hamiltonian\footnote{The following notation is adopted in this paper: $\vec{H}$ denotes a vector Hamiltonian and $H$ denotes energy.} of the form

\begin{equation}
\vec{H} = c \; (\vec{p} \: [\alpha]) + m_0 c^2 \: \vec{\xi}
\label{i1}
\end{equation}

\vspace*{.08in}

where $c$ is the velocity of light, $m_0$ is the rest mass of the particle,  $[\alpha]$ is a 4$\times$4 matrix, and $\vec{\xi}$ and $\vec{p}$ are 1$\times$4 vectors, $\vec{p}$ being the momentum vector. To check his theory, Dirac proceeded to calculate the $x$ component of the velocity of the particle, $\dot{x}$, from the Poisson bracket

\begin{equation}
\dot{x} = [x, \vec{H}] = \frac{\partial \vec{H}}{\partial p_x} = c \; \alpha_{11}.
\label{i2}
\end{equation} 

\vspace*{.08in}

Since $\alpha_{11} = \pm 1$, he concluded that $\dot{x}$ must be always equal to $\pm c$. Upon such conclusion, he wrote: ``This result is rather surprising ...since electrons are observed in practice to have velocities considerably less than that of light, it would seem that we have here a contradiction with experiment". He then went on to conclude that: ``The contradiction is not real, though, since the theoretical velocity in the above conclusion is the velocity at one instant of time while observed velocities are always average velocities through appreciable time intervals". So Dirac chose to ignore that serious contradiction between theory and experiment and offered instead an interpretation that does not correlate with physical reality and which is still, to date, not supported by any experimental evidence. Nevertheless, the soundness of Dirac's theoretical approach allowed him to make correct predictions, such as about the existence of spin and of antiparticles. Beside the early difficulties of the beta-decay and of Dirac's equation, a number of other difficulties emerged later. Most notably, the fact that the measured yield of the nuclear explosions that were carried during the 1940s was found to be substantially less than the value of the energy release expected theoretically\cite{Serber}. The only explanation that this fact was attributed to was that the rapid disintegration of the mechanical structure surrounding the uranium or plutonium core prevents the chain reaction from getting completed (though admittedly the entire reaction takes less than $10^{-6}$ sec. to complete). In this paper a result will be shown, however, that very strongly suggests that the true reason was likely theoretical rather than mechanical. Another problem that clearly emerged over the years was the fact that experiments which are designed to measure the masses of various particles usually show a large discrepancy between the experimental value and the theoretically expected value of the mass of a given particle if the experiment is carried at low energies. If, however, the experiment is carried at very high energies, then the experimentally determined value of the mass is usually found to be in very good agreement with the theoretical value. While the list is extensive, we can mention the meson experiments\cite{Perkins2},\cite{Lattes},\cite{Bjorklund} as examples of experiments that were carried at low energies and in which that fact was apparent. As examples of experiments that were carried at high energies we refer to the detection of the $W$ and $Z$ bosons in the early 1980s \cite{CERN1},\cite{CERN2}.\\
\\
In this paper we will take a long step back in time and attempt to determine the fundamental cause of all the various difficulties that we discussed. It will be demonstrated, beyond any doubt, that the cause is the fundamental inconsistency of the relativistic mass-energy equivalence principle that was introduced by Einstein in 1905 with the basic assumptions of the wave mechanics introduced by Compton and de Broglie 18 years later. This paper is organized as follows: in Sec.2 it will be demonstrated that the Compton-de Broglie wave mechanics has its own, very specific view, of the principle of mass-energy equivalence. In Sec.3, the difference between the wave-mechanical view and the relativistic view of  mass-energy equivalence will be explained. In Sec.4 we will demonstrate, directly from Einstein's original papers on special relativity\cite{Einstein1},\cite{Einstein2},\cite{Einstein3}, that there are two different and distinct ways to interpret the relativistic concept of mass-energy equivalence. One of the interpretations leads to a result that is in total agreement with wave mechanics, while the other interpretation leads to a result that is in total disagreement. Unfortunately, the interpretation and hence the result that is in disagreement with wave mechanics is the one that was selected by Einstein in 1905 (and remained as a standard since then). By selecting the interpretation that does lead to an agreement with wave mechanics we will be able to solve, as examples, two of the substantial problems that we highlighted above. In Sec.5, we will rework Dirac's theory of the relativistic wave equation of a free particle. Unlike Dirac's result, the result obtained here does agree with experimental observation. In Sec.6, the problem of the large discrepancy between the theoretical yield and the actual yield from nuclear fission will be treated. It will be shown that the modified view introduced here is in fact in very good agreement with the published experimental data. Sec.7 contains a comment on the nature of beta-decay and the problem of the shortfall in the solar neutrinos. Sec.8 summarizes the conclusions of this paper.\\
\\
\\
{\large\bf 2. The wave-mechanical view of the principle 
of mass-energy equivalence:}\\
\\
The fundamental postulate of Compton and de Broglie,

\begin{equation}
\lambda = \frac{h}{p},
\label{Compton}
\end{equation}

which relates the wavelength $\lambda$ to the momentum $p$ of an elementary particle, if taken together with the mathematical representation of de Broglie's matter waves, namely, the traveling wave equation, lead to a simple and important result that is in direct contradiction with the relativistic principle of mass-energy equivalence. Although de Broglie and other researchers had offered a number of proposals and suggestions that seem, superficially, to resolve the contradiction, careful examination will show that the contradiction was never actually resolved. We shall now proceed to make a number of important observations, ending with the result that is inconsistent with relativity. We will then demonstrate that this result represents a very specific view of the principle of mass-energy equivalence; a view that is inherently incompatible with the relativistic view. We will attempt to keep the mathematical developments at a minimum so that the facts can be clearly seen.\\
\\
If we take a one-dimensional de Broglie wave function\cite{Dirac}

\begin{equation}
\psi (x, t) = A \exp i(k x - \omega t)
\label{1}
\end{equation}

where $A$ is the amplitude constant, $k = 2\pi / \lambda$ is the propagation constant, and $\omega = 2\pi\nu$ is the angular frequency (where $\nu$ is the frequency), we then have, by definition, $\omega=k\lambda\nu$. Now, since the product $\lambda\nu$ is the propagation velocity (or phase velocity), $u$, of the wave, then $\omega=k u$. Using this relationship in Eq.(\ref{1}) gives

\begin{equation}
\psi (x, t) = A \exp ik(x - u t).
\label{3}
\end{equation}

From Eq.(\ref{3}) we get

\begin{equation}
\frac{\partial^2 \psi}{\partial x^2} = - k^2 \psi
\label{4}
\end{equation}

In the special case where the propagation velocity $u$ is equal to $c$, we get an interesting result. Since $\omega=k c$, and since $c^2 = 1 / \epsilon_0 \mu_0$,
where $\epsilon_0$ and $\mu_0$ are the permittivity and permeability of free space, respectively, then $k^2=\omega^2 \epsilon_0 \mu_0$. By substitution into Eq.(\ref{4}) we obtain

\begin{equation}
\frac{\partial^2 \psi}{\partial x^2} = - \omega^2\epsilon_0\mu_0 \; \psi.
\label{7}
\end{equation}

We should recognize that Eq.(\ref{7}) is the Helmholtz equation. It can be equally derived directly from Maxwell's equations\cite{Hayt}, with the result that the wave function $\psi$ is actually the electric field intensity $E$. Through similar reasoning, it was concluded previously\cite{Dirac} that the de Broglie matter wave is indeed an electromagnetic wave when $u=c$. Moving back to the general case (i.e., $u\neq c$), we can further see from Eq.(\ref{3}) that

\begin{equation}
\frac{\partial\psi}{\partial t} = - i k u \; \psi = - i \omega \; \psi.
\label{9}
\end{equation}

Multiplying the above expression by $i \hbar$ yields 

\begin{equation}
i \hbar \frac{\partial\psi}{\partial t} = \hbar\omega \; \psi = \hbar (2\pi\nu) \; \psi = h \nu \; \psi.
\label{10}
\end{equation}

Since Eq.(\ref{10}) is the generally accepted form of the non-rela\-tivistic Schr\"{o}dinger equation, then the multiplicand of $\psi$ on the right-hand side must be the total energy $H$ of the particle; that is, the total energy is $H = h \nu$ (this result confirms de Broglie's postulate about the total energy of the particle).  Now since

\begin{equation}
\lambda = \frac{h}{p} = \frac{u}{\nu},
\label{12}
\end{equation}

we see that

\begin{equation}
H = h \nu = p u.
\label{13}
\end{equation}

For de Broglie's matter wave hypothesis to be valid we must conclude, and indeed it has been demonstrated\cite{Dirac},\cite{Eisberg}, that the propagation velocity $u$ of the wave is equal to the particle's velocity $v$ (one important concept, originally introduced by de Broglie and later promoted by others, is that a matter wave is actually a wave packet, or wave group, that results from the superposition of many individual waves. However, since Eq.(\ref{13}) is a total energy equation, then we have no alternative but to conclude that the wave in question is the overall wave packet -if that concept is in fact valid- and hence $u=v$). Now since $p=mv$, where $m$ is the particle's mass, then Eq.(\ref{13}) becomes 

\begin{equation}
H = h \nu = m v^2.
\label{14}
\end{equation}

Once again we note that in the special case of $v=c$, the well-known relativistic mass-energy transformation law $H = m c^2$ is obtained. Accordingly, from a strictly wave-mechanical point of view, the relativistic expression of mass-energy equivalence emerges as a special case only.\\
\\
\\
{\large\bf 3. Disagreement of the Compton - de Broglie mechanics with relativity:}\\
\\
We have shown that the Helmholtz wave equation and Einstein's mass-energy transformation law both emerge as a special case from the Compton-de Broglie wave mechanics, that is, when the velocity $v$ of the particle and its associated wave is equal to $c$. At velocities other than $c$ (i.e., at sub-relativistic velocities), the expression in Eq.(\ref{14}) is obviously in disagreement with relativity, since relativity would maintain that the total energy must still be expressed as $m c^2$. We must now point out that some authors\cite{Wehr} have suggested in the past that by taking the total energy to be $H=pu=mvu=mc^2$, then the velocity of the matter wave will be given by $u = c^2 /v$; hence $u > c$. Such result is clearly an impossible result, however, given the emergence of Helmholtz's equation, and given the above observation about Eq.(\ref{13}) (i.e., the fact that $u$ must be equal to $v$). Simply stated, the total energy of the particle, as defined by the fundamental postulates in Eqs.(\ref{12}) and (\ref{13}), leads to inconsistencies with the theory of special relativity which are unsolvable (even if we accept such concepts as the concept of the wave packet). We also note here that if we substitute for $m$ in Eq.(\ref{14}) by its relativistic expression 

\begin{equation}
m = \frac{m_0}{\sqrt{1- v^2 / c^2}},
\label{15}
\end{equation}

this of course still does not change the fundamental disagreement between the wave-mechanical and the relativistic energy expressions (that is, $m v^2$ versus $m c^2$. Note that the total energy of the particle at rest, or $H_0$, is equal to $m_0 c^2$ in special relativity, while it is precisely zero in wave mechanics).\\
\\
In summary, we can clearly see that the wave-mechanical prediction of the total energy of a particle is in serious and fundamental contradiction with the relativistic prediction, and the limited efforts that were carried in the past  to reconcile the two views have not had any success.\\
\\
{\large\bf 4. Origin of the law $H = m c^2$ and an alternative understanding of the relativistic concept of mass-energy equivalence:}\\
\\
We shall now proceed to investigate the origin of the mass-energy equivalence principle directly from Einstein's original papers on relativity\cite{Einstein1},\cite{Einstein2},\cite{Einstein3} and attempt to develop an understanding of the reason for the disagreement between Einstein's kinematics and the wave mechanics of Compton and de Broglie. As we shall see, the two theories can in fact be reconciled perfectly, though such reconciliation will be in a manner that was not intended or envisioned by Einstein.\\
\\
{\bf 4.1. The basic observation:}\\
\\
In the first relativity paper, Einstein considered the commonly applied form of Newton's second law of motion, $F = m \, a$. To express that law within the framework of relativity, he obtained a relativistic transformation term for the mass given by $m = m_0/(1-v^2/c^2)^\frac{3}{2}$ (which he labeled ``longitudinal mass"). He then carried the integration of $F$ over a distance $X$ to get an expression for the work done on the particle, which must be equal to the kinetic energy $E_k$ of the particle. The result of the integration was

\begin{equation}
E_k = m_0 c^2 \; \left( \frac{1}{\sqrt{1-v^2/c^2}} - 1 \right).
\label{16}
\end{equation}

From that expression, Einstein (and others) concluded that\cite{Einstein3} 

\begin{equation}
\mbox{Total particle energy} = m c^2 = E_k + m_0 c^2 
\label{17}
\end{equation}

Eq.(\ref{17}) has been the standard equation on which virtually all modern post-relativistic particle physics is based. In addition to the total energy expressed as $m c^2$, Eq.(\ref{17}) also predicts the existence of a ``rest energy" that is given by the term $m_0 c^2$. Originally, it was not clear from Einstein's derivation how the mass multiplier term $1/(1-v^2/c^2)^\frac{3}{2}$ was obtained. Later, however, a number of authors\cite{Born},\cite{French} pointed out that most likely Einstein's approach was the following: since Newton's second law can be more fundamentally expressed as ``Force = rate of change of momentum", or $F= dp/dt$, and since the momentum can be expressed as

\begin{equation}
p = mv = \frac{m_0}{\sqrt{1- v^2 / c^2}} \; v,
\label{17b}
\end{equation}

then the kinetic energy of the particle can be obtained from

\begin{equation}
E_k = \int_0^X F \; dx = \int_0^X \frac{dp}{dt} \; dx.
\label{19}
\end{equation}

The integral of Eq.(\ref{19}) can obviously be written in the alternative form

\begin{equation}
E_k = \int_0^p v \; dp
\label{19b}
\end{equation}

There are two ways for calculating this integral. We shall first describe a method that leads to Einstein's equation. If we re-write the integral as

\begin{equation}
E_k = \int_0^p v \; dp = \int_0^v v \; (\frac{dp}{dv}) \; dv,
\label{19c}
\end{equation}

and use the easily-verifiable expression

\begin{equation}
\frac{dp}{dv} = 
\frac{m_0}{(1-v^2/c^2)^{\frac{3}{2}}},
\label{19d}
\end{equation} 

then we have 

\begin{equation}
E_k = \int_0^v 
\frac{m_0 v}{(1-v^2/c^2)^{\frac{3}{2}}} \; dv
\label{19e}
\end{equation}

This is precisely the expression that appeared in Einstein's first paper\cite{Einstein1} and which led to the formula given by Eq.(\ref{16}). We shall now describe the second method for calculating the integral of Eq.(\ref{19b}). Although the result must be the same numerically, we shall see that the physical implications of the result are potentially different. If we integrate by parts, Eq.(\ref{19b}) is re-written as

\begin{equation}
E_k = \int_0^p v \; dp = pv - \int_0^v p \; dv,
\label{20}
\end{equation}

or,

\begin{equation}
E_k = m v^2 - \int_0^v  \frac{m_0 v}{\sqrt{1 - v^2/c^2}} \; dv
\label{21}
\end{equation}

Carrying the integration in Eq.(\ref{21}) results in

\begin{equation}
E_k = m v^2 - m_0 c^2 (1 - \sqrt{1 - v^2/c^2}) 
\label{22}
\end{equation}

It is not difficult to verify that equations (\ref{22}) and (\ref{16}) are the same, that is

\begin{eqnarray}
E_k & = & m v^2 - m_0 c^2 (1 - \sqrt{1 - v^2/c^2}) \nonumber\\
    & = & m_0 c^2 \; \left( \frac{1}{\sqrt{1-v^2/c^2}} - 1 \right);
\label{23}
\end{eqnarray}

(this can be verified by substituting for $m$ by its relativistic expression and rearranging the terms). Now, in light of the conclusion about the total energy obtained in the previous section, that is, if indeed we can express the total energy of the particle by the quantity $m v^2$, then Eq.(\ref{22}) would lead us to conclude that

\begin{eqnarray}
\lefteqn{\mbox{Total particle energy} = m v^2 = } \qquad\qquad  \nonumber\\
  & & E_k + m_0 c^2 (1 - \sqrt{1 - v^2/c^2}),  \nonumber\\ 
  & & \label{24}
\end{eqnarray}

which is of course significantly different from Einstein's expression in Eq.(\ref{17}). The relevant difference between the two expressions is, of course, that Eq.(\ref{24}) agrees with the wave-mechanical prediction while Einstein's expression does not. The obvious question now is: which expression is ultimately correct? To answer that question, we note the following very important fact: Einstein derived the kinetic energy equation first and then offered the mass-energy transformation law of Eq.(\ref{17}) as an interpretation\cite{Einstein2},\cite{Einstein3}. There was no direct derivation of that law from the fundamental axioms of relativity, as can be clearly concluded from those two subsequent papers (see additional remarks in Sec.4.2 below). The mass-energy transformation law was merely an interpretation of the original kinetic energy equation $E_k = m c^2 - m_0 c^2$. Had Einstein carried the integration differently -as we have just shown above- he would have obtained the expanded form of that equation, that is,

\begin{eqnarray}
E_k & = & m v^2 - m_0 c^2 (1 - \sqrt{1 - v^2/c^2}), \label{24b}\\ 
    &   & (= m c^2 - m_0 c^2) \nonumber
\end{eqnarray}

and in all likelihood he would have concluded that the total energy of the particle must be given by the quantity $m v^2$. In other words, we can clearly see that the wave-mechanical total energy expression $H = m v^2$ is indeed distinguishable from Einstein's kinematics, while the expression $H = m c^2$ was essentially the result of subconscious interpretation of the original kinetic energy equation, shown between parenthesis under Eq.(\ref{24b}) above. Of course, Einstein was a predecessor to de Broglie and had no guideline to follow when he made that  original interpretation. In view of these facts, it is clear that the theory of relativity should have been reconciled with wave mechanics by taking the expression $H = m v^2$ as the total energy formula. It is unfortunate, however, that such a step was not taken for so many years, especially in view of the large number of theoretical results that do not seem to correlate with experimental data.\\
\\
If we now assume that Eq.(\ref{24}) is valid, then we must conclude that the term $m_0 c^2 (1 - \sqrt{1 - v^2/c^2})$ is the mass-energy term. We must further conclude that the mass-energy is velocity-dependent and hence is equal to zero at rest. An interesting test of Eq.(\ref{22}) will be to expand the term $\sqrt{1 - v^2/c^2}$ in the equation into a power series, obtaining

\begin{equation}
\sqrt{1 - v^2/c^2} = 1 - \frac{1}{2} \frac{v^2}{c^2} - \frac{1}{8} \frac{v^4}{c^4} - \ldots
\label{24c}
\end{equation}

Eq.(\ref{22}) therefore becomes

\begin{equation}
E_k = m v^2 - m_0 c^2 \; (\frac{1}{2} \frac{v^2}{c^2} + \frac{1}{8} \frac{v^4}{c^4} + \ldots )
\label{25}
\end{equation}

For small velocities, that is, $v\ll c$, we can see that Eq.(\ref{25}) gives

\begin{equation}
E_k \approx m_0 v^2 - \frac{1}{2} m_0 v^2 = \frac{1}{2} m_0 v^2,
\label{25b}
\end{equation}

which is the classical expression of kinetic energy. Of course, we can further make the conclusion here that the mass-energy term is approximately equal to $1/2 \: m_0 v^2$ at small velocities. This last observation is important and will be used again in Sec.6. (Note that the dependence of the mass-energy quantity on the velocity is not in contradiction with the principle of  conservation of energy, since, in relativity, energy is a relative quantity, along with other physical quantities like time and mass. See for example the dependence of the observed energy on the velocity in the treatment of the Doppler effect in the first relativity paper\cite{Einstein1}).\\
\\
{\bf 4.2. Remarks about references [2],[3] and the law of photonic exchange:}\\
\\
In the second relativity paper\cite{Einstein2}, Einstein stated the law $H=mc^2$ for the first time as a ``general conclusion" from the kinetic energy equation $E_k = m c^2 - m_0 c^2$ (after deriving a law of radiation that is discussed below). Perhaps realizing that this general conclusion was not clearly explained, he elaborated on it further in a third paper\cite{Einstein3} that was published in 1907. In that paper, Einstein wrote the expression: Total Energy =  Kinetic-Energy + Mass-Energy. He indicated (without any proof) that the mass-energy quantity must be a constant. He then referred again to the kinetic energy equation $E_k = m c^2 - m_0 c^2$ and wrote:\\
\\
``One then obtains:\\
$\mbox{Total Energy} = m_0 c^2 / \sqrt{1 - v^2/c^2}$"\\
\\
(note that the modern notation is used here instead of the actual notation shown in the paper). As we indicated earlier, such interpretation of the kinetic energy equation has remained unquestioned to this day.\\
\\
Concerning the law of radiation that was shown in the second paper (which we can refer to in modern terms as the law of photonic exchange), we must note some important facts. In that paper, Einstein considered the equivalence between a quantum of radiation $L$ and the kinetic energy $E_k$ of a particle. First we must observe, in view of the above discussion, that the kinetic energy {\em is} numerically equal to $\Delta m c^2$. If a particle emits a quantum of radiation and its kinetic energy is altered, then the energy contained in such a quantum will be numerically equal to $\Delta m c^2$ (where $\Delta m$ is the ``apparent'', or relativistic mass change). Einstein did correctly reach this conclusion (though he used a series of very primitive assumptions in his analysis and indicated that the result can be obtained by expanding the Lorentz transformation terms, which was an unnecessary step). The problem now lies with the ``general conclusion'' that was stated after that derivation. As we have demonstrated above, the total energy of the particle {\em is not} equal to $m c^2$, but rather $m v^2$. If the particle is totally annihilated and a pair of photons is emitted, then the energy contained in such a pair will be numerically equal to $m v^2$, where $m$ is the total relativistic mass of the particle. In other words, there is a very clear and strong distinction between the kinetic energy and the total energy of a particle, and while the former is numerically equal to $\Delta m c^2$, the latter is numerically equal to $m v^2$. Such distinction was not made by the standard formulation of special relativity.\\
\\
We must finally point out that the equation $H=mv^2$ leads to a very important conclusion: obviously, mass-energy transformations will not be possible unless relative motion is present. If a particle decays, for instance, then the decay energy will not be observed by an observer who is stationary with respect to that particle; it will be observed only by an observer who is in relative motion with respect to that particle (indeed, some experiments\cite{Steinberger} that sought in the past to detect the decay of stationary particles have showed negative results). In addition, we must conclude that photonic exchanges in general are not possible unless relative motion is present (since $H=0$ if $v=0$). This last conclusion, of course, does agree with our current understandings from the theory of Quantum Electrodynamics (QED).\\
\\
To summarize the above discussion: had Einstein obtained the kinetic energy expression $E_k = mv^2 - m_0c^2 (1-\sqrt{1 - v^2/c^2})$ instead of the original expression $E_k = mc^2 - m_0 c^2$, he would have made the conclusions that we reached above instead of the original conclusions that resulted as standard from the theory of special relativity.\\
\\
{\bf 4.3. Special illustrative example: the process of electron-positron annihilation:}\\
\\
A particular case to consider in view of the above conclusions is the annihilation of an electron and a positron at low energy, in which two photons of 0.511Mev each are emitted. The current view about this process is that the two particles annihilate each other strictly at rest, and hence the quantity $m_0c^2$ is obtained. Given the conclusions that we just reached, however, we must conclude that the particles cannot annihilate at rest. They will have to exchange virtual photons and hence accelerate toward each other before annihilation. As the particles accelerate toward each other they will gain kinetic energy but lose potential energy (i.e., they gain a potential energy term with a negative sign, since they will be bound by the electrostatic force). The total energy of each particle will be therefore given by: Total Energy = $mv^2$ + Potential Energy = Kinetic Energy + Potential Energy + Mass Energy (this defines the total Hamiltonian of the particle). If the amount of external work exerted on the particles is negligible, then the kinetic and the potential energy terms will precisely cancel each other and hence the total energy will be approximately equal to the mass energy, or the quantity $m_0 c^2 (1 - \sqrt{1 - v^2/c^2})$. If the particles reach a velocity $v \approx c$ before annihilation, then the total energy for each particle will be approximately equal to $m_0c^2$. By adding a small, variable amount of external work to the energies of the two particles (which may be thermal or electrostatic work), then the total energy will have a Gaussian distribution around the mean value $m_0c^2$. This distribution has been observed experimentally and is currently known as the ``Doppler broadening" of the 0.511Mev line (though such interpretation is obviously inconsistent with the fact that photon emission occurs {\em after}, not before, annihilation).\\
\\
This short description explains the process of electron-positron annihilation in view of the conclusions reached above. Clearly, the quantity $m_0c^2$ is obtained in this process as a special case. As we shall demonstrate later, if the velocities of particles which are involved in various other interactions are not close to $c$, then this special case is not obtained; therefore it cannot be generalized to encompass all aspects of particle physics (which is the current assumption).\\
\\
\\
{\large\bf 5. Modified relativistic theory for the motion of a free particle:}\\
\\
In view of the foregoing, we shall now assume that the correct total energy expression is $H = m v^2$ and examine what the implications are for Dirac's free-particle theory. As we indicated in Sec.1, Dirac obtained a Hamiltonian that led him to conclude that the velocity of any particle must be always equal to $\pm c$, which is of course a result that does not have any correlation with physical reality. Although Dirac speculated\cite{Dirac} at that time that the velocity might actually be composed of a steady-state part and an oscillatory part, such speculation was not confirmed experimentally. Moreover, Dirac attempted but was unable to show mathematically how the proposed two components would add up to be precisely equal to $c$. We shall now proceed to rework Dirac's theory, using $H= m v^2$, and obtaining a new  Hamiltonian in the process. As we shall verify, the new Hamiltonian will  directly lead to the result that the velocity must be equal to $\pm v$; which is of course a result that is in sharp contrast with Dirac's result and which does agree with experimental observation. We shall also verify that the spin magnetic moment term obtained by Dirac remains unchanged in the present formulation.\\
\\
{\bf 5.1. The wave equation:}\\
\\
We shall begin by describing briefly Dirac's approach for obtaining the relativistic wave equation and then proceed to derive the modified equation and hence the modified Hamiltonian. Dirac considered the mass of the particle as represented by its relativistic expression $m = m_0 / \sqrt{1-v^2/c^2}$. If we square that expression and rearrange the terms, we get

\begin{equation}
m^2 c^2 = m^2 v^2 + m_0^2 c^2.
\label{26}
\end{equation}

Multiplying by $c^2$, we get

\begin{equation}
m^2 c^4 = m^2 v^2 c^2 + m_0^2 c^4.
\label{27}
\end{equation}

But since $m c^2$ is the total energy according to Einstein, then we have

\begin{equation}
H^2 = p^2 c^2 + m_0^2 c^4,
\label{28}
\end{equation}

or

\begin{equation}
H = c \; \sqrt{p^2 + m_0^2 c^2}.
\label{29}
\end{equation}

Since the term $p^2$ can be written as $\sum_r p_r^2$, where $p_r$ is a one-dimensional momentum component and $r=1,2,3$, we finally have

\begin{equation}
H = c \; \sqrt{\sum_r p_r^2 + m_0^2 c^2}.
\label{30}
\end{equation}

This was Dirac's total energy equation and was subsequently used to obtain the relativistic wave equation. To obtain the modified wave equation, we now proceed to multiply Eq.(\ref{26}) by $v^2$, getting

\begin{equation}
m^2 c^2 v^2 = m^2 v^4 + m_0^2 c^2 v^2.
\label{31}
\end{equation}

Using $H= m v^2$ as the total energy of the particle, we have, from the above expression

\begin{equation}
H^2 = p^2 c^2 - m_0^2 c^2 v^2. 
\label{32}
\end{equation}

Now, since $v^2 = \sum_r v_r^2$, where $v_r$ is a one-dimensional velocity component, Eq.(\ref{32}) can equivalently be written as

\begin{equation}
H = c \; \sqrt{\sum_r p_r^2 - m_0^2 \sum_r v_r^2}.
\label{33}
\end{equation}

Eq.(\ref{33}) can be further simplified by noting that $v_r = p_r /m$. We finally have

\begin{equation}
H = c \; \sqrt{(1 - m_0^2 / m^2) \sum_r p_r^2}.
\label{34}
\end{equation}

Following Dirac's approach, if we let $\vec{p_0}$ be a vector defined as  $\vec{p_0} = \vec{H} /c$, where $\vec{H}$ may be Hamiltonian of the form $\vec{H} = (H,0,0)$, we will seek a wave equation that is linear in $\vec{p_0}$. If we take an equation of the most simple, basic form

\begin{equation}
(\vec{p_0} - \sum_r \vec{p_r} \: [\alpha_r]) \psi = 0,
\label{35}
\end{equation}

this form can be sufficient without any additional terms if we do not impose any restrictions on the matrices $[\alpha_r]$. Dirac found that such matrices must be non-commuting, but it is obvious here that such matrices must also contain mass terms. Multiplying Eq.(\ref{35}) by the vector $(\vec{p_0} + \sum_r \vec{p_r} \: [\alpha_r])$, we get

\begin{equation}
p_0^2 - (\sum_r \vec{p_r} \: [\alpha_r])^2 = 0
\label{36}
\end{equation}
  
Comparing this last expression with Eq.(\ref{34}), we conclude that


\begin{eqnarray}
\lefteqn{(1 - m_0^2 / m^2) \sum_r p_r^2 = (\sum_r \vec{p_r} \: [\alpha_r])^2 =} \nonumber\\ 
 &  & \sum_r \vec{p_r} \: [\alpha_r]^2 \: \vec{p_r}^T 
 + \sum_j \sum_k \vec{p_j} \: [\alpha_j] [\alpha_k] \: \vec{p_k}^T, \nonumber\\
 & &
\label{37}
\end{eqnarray}

where $j,k=1,2,3$, and $j \neq k$. Accordingly, the matrices $[\alpha_r]$ must satisfy

\begin{equation}
[\alpha_r] = \pm \sqrt{1 - m_0^2 / m^2} \; [\beta_r],
\label{38}
\end{equation}

where $[\beta_r]$ are matrices that must satisfy the two conditions:

\begin{equation}
[\beta_r]^2 = I, \qquad \mbox{and} \qquad 
[\beta_j] [\beta_k] + [\beta_k] [\beta_j] = 0
\label{39}
\end{equation}

Examples of such matrices were suggested by Dirac\cite{Dirac}. They might take the following forms, among others,

\begin{equation}
\left( \begin{array}{cccc}
0 & 0 & 0 & 1\\
0 & 0 & 1 & 0\\
0 & 1 & 0 & 0\\
1 & 0 & 0 & 0
\end{array} \right), \quad
\left( \begin{array}{cccc}
0 & 0 & 0 & -i\\
0 & 0 & i & 0\\
0 & -i & 0 & 0\\
i & 0 & 0 & 0
\end{array} \right).
\label{40}
\end{equation}

(Note that Dirac used $4 \times 4$ matrices to account for time as the fourth dimension. It was independently confirmed later that the minimum number of dimensions that will satisfy Dirac's theory is in fact four).\\
\\
Using the relativistic expression for $m$, the matrices $[\alpha_r]$ can now be written as

\begin{equation}
[\alpha_r] = \pm \sqrt{1 - (1- v^2/c^2)} \; [\beta_r] = 
\pm \frac{v}{c} \: [\beta_r];
\label{41}
\end{equation}

and therefore, from (\ref{37}) and (\ref{41}), the vector Hamiltonian can be written as

\begin{equation}
\vec{H} = c \: \vec{p_0} = c \; \sum_r \vec{p_r} \: [\alpha_r] =
\pm v \; \sum_r \vec{p_r} \: [\beta_r]
\label{42}
\end{equation}

To check the modified theory, it can be now easily verified that the velocity component $\dot{x_1}$ will be given by

\begin{equation}
\dot{x_1} = [x_1, \vec{H}] = \pm v.
\label{43}
\end{equation}

Unlike Dirac's result, this result is of course in agreement with experimental observation. It is important to note here that, mathematically, $\dot{x_1}$ is the ``expected'' value of the velocity. From Eq.(\ref{42}), we can also see that the negative energy states are still preserved here.\\
\\
{\bf 5.2. Motion of a charged particle in a magnetic field:}\\
\\
We now consider the motion of a charged particle in a magnetic field to obtain a formulation for the spin magnetic moment term that must appear in the final Hamiltonian (we assume the absence of an electrostatic field here). In the presence of a magnetic field, the change in the particle's momentum $\Delta p$ that occurs as a result of the interaction with the field is given by\cite{Feynman}

\begin{equation}
\Delta p = \frac{e}{c} \: A,
\label{44}
\end{equation}

where $e$ is the particle's charge and $A$ is the magnitude of the vector magnetic potential. Adding that term to the momentum in Eq.(\ref{42}) gives the Hamiltonian

\begin{equation}
\vec{H} = \pm v \; \sum_r (\vec{p_r} + \frac{e}{c} \: \vec{A_r}) \: [\beta_r]
\label{45}
\end{equation}

By squaring Eq.(\ref{45}), we get

\begin{eqnarray}
\lefteqn{\frac{H^2}{v^2}  =  \; \sum_r \left[ 
(\vec{p_r} + \frac{e}{c} \: \vec{A_r}) \: [\beta_r] \right]^2 } \quad \nonumber\\
 & & + \sum_j \sum_k (\vec{p_j} + \frac{e}{c} \: \vec{A_j}) [\beta_j][\beta_k]
(\vec{p_k} + \frac{e}{c} \: \vec{A_k})^T \nonumber\\
 & & \label{46}
\end{eqnarray} 

It is fairly easy to verify that the second term on the r.h.s. of the above expression must vanish since the $\vec{p_r}$ vectors commute and since the $[\beta_r]$ matrices satisfy condition (\ref{39}). In Dirac's treatment of the subject, he was able to show that the following equation holds

\begin{equation}
\left[ 
(\vec{p_r} + \frac{e}{c} \: \vec{A_r}) \: [\beta_r] \right]^2 =
(\vec{p_r} + \frac{e}{c} \: \vec{A_r})^2 + \frac{\hbar e}{c} \:
\| \vec{M} \: [\beta_r] \|,
\label{47}
\end{equation}

where $\vec{M} = curl \vec{A}$ is the magnetic field intensity vector. Eq.(\ref{46}) therefore becomes

\begin{equation}
H^2 = v^2 \; \sum_r (\vec{p_r} + \frac{e}{c} \: \vec{A_r})^2 
+ v^2 \; \frac{\hbar e}{c} \: \sum_r \: \| \vec{M} \: [\beta_r] \|.
\label{48}
\end{equation}

(Note that Eq.(\ref{48}) is a scalar equation). If we now let $H = m v^2$ and divide both sides of the equation by $2m v^2$, we get

\begin{equation}
\frac{1}{2} \: mv^2 = \frac{1}{2m} \sum_r (\vec{p_r} + \frac{e}{c} \: \vec{A_r})^2 
+ \frac{\hbar e}{2mc} \: \sum_r \| \vec{M} \: [\beta_r] \|
\label{49}
\end{equation}

If the particle is an electron, then $e$ is a negative quantity and the above equation becomes

\begin{equation}
\frac{1}{2m} \sum_r (\vec{p_r} + \frac{e}{c} \: \vec{A_r})^2 = \frac{1}{2} \: mv^2 +
\frac{\hbar |e|}{2mc} \: \sum_r \| \vec{M} \: [\beta_r] \|
\label{50}
\end{equation}

Without the presence of the magnetic field, the l.h.s. of Eq.(\ref{50}) is reduced to

\begin{equation}
\frac{1}{2m} \sum_r p_r^2 = \frac{p^2}{2m},
\label{51}
\end{equation}

which is the same as $1/2 \: mv^2$. We can therefore conclude that the second term on the r.h.s. of Eq.(\ref{50}) is the term that represents the interaction of the field with the electron's magnetic moment. Hence the quantity $\hbar |e| / 2mc$ is the spin magnetic moment coefficient. In general, we can withdraw here the following two important conclusions: 1. The modified theory fully yielded the classical expression of kinetic energy, with the addition of the spin interaction term; and 2. The spin interaction term obtained here is the same as the one obtained by Dirac\cite{Dirac} (which is one Bohr magneton). The second conclusion is a confirmation that this part of Dirac's theory was correct. The first conclusion, however, shows a fact that was not apparent from Dirac's theory. Specifically, when $\vec{M}=0$ (that is, when the particle is away from the magnetic field lines), Eq.(\ref{50}) becomes

\begin{equation}
\frac{1}{2m} \sum_r (\vec{p_r} + \frac{e}{c} \: \vec{A_r})^2 = \frac{1}{2} \: mv^2,
\label{51b}
\end{equation}

which is a direct confirmation of the Aharonov-Bohm effect\cite{Aharonov},\cite{Imry}. Clearly, Eq.(\ref{51b}) shows that the components $p_r$ of the momentum will be altered while the kinetic energy remains constant.\\
\\
\\
{\large\bf 6. The problem of the energy released from nuclear fission:}\\
\\
\underline{Note:} The information contained in this section is based on previously known public information.\\
\\
As we indicated previously in Sec.1, it is currently believed that the efficiency of nuclear fission as an energy source is extremely low. The efficiency is defined as the ratio of the amount of energy (or yield) obtained from a given mass of fissionable material to the theoretical value of the total energy that is expected to be released from that mass. In nuclear weapons as well as in nuclear reactors, that efficiency figure is typically on the order of a few percentage points (though efficiencies as high as 40\% have been achieved). Usually, it is assumed that most of the atoms of the fissionable material simply do not undergo fission in the chain reaction and are thus ``wasted'' in the process. As a specific case to consider, Serber\cite{Serber} indicated that the efficiency of the ``Little Boy'' weapon that was used against Hiroshima in 1945 was about 2\%. In this section, we shall rework the calculations given by Serber for that specific example, using the equation $H = m v^2$. As we shall see, the efficiency will actually prove to be much higher; but more importantly, we will get a theoretical estimate for the energy release that is in very good agreement with the actual measured value.\\
\\
{\bf 6.1. Energy of fission as a function of the nuclear excitation:}\\
\\
It has been known for at least three decades that the equation $H=mc^2$ does not give an accurate estimate of the total kinetic energy of the fragments that result from the fission of a nucleus. In laboratory experiments using thermal neutrons\cite{Hambsch},\cite{Thierens}, for instance, it is usually found that the total kinetic energy (TKE) of the fragments that result from the fission of either $U^{235}$ or $Pu^{239}$ is 20-60Mev less than the Q-value of the reaction predicted by the quantity $\Delta m c^2$. It is typically assumed that the difference between the Q-value and the TKE of the fragments is lost into unobservable effects, such as additional excitation energy of the fragments. As we indicated above, that energy deficit can be as high as 98\% in nuclear weapons (though the usual assumption in that case is that the chain reaction simply does not complete). Our objective now is to calculate a Q-value for the fission reaction of $U^{235}$ by using the quantity $\Delta m v^2$ instead of the quantity $\Delta m c^2$. As we shall verify, the TKE will prove to be in very good agreement with the numerical value of the quantity $\Delta m v^2$. We shall also reach the conclusion that the quantity $\Delta m c^2$ represents a maximum theoretical limit for the energy that can be obtained from nuclear fission.\\
\\
Recent models\cite{Straede},\cite{Wilkins} of the nuclear fission process have sought to obtain a precise numerical prediction of the TKE of the fragments by extending the successful liquid-drop model originally introduced by Bohr and Wheeler. In those models, the equation that is relied upon for predicting the TKE is the Coulomb potential energy equation, that is, 

\begin{equation}
\mbox{TKE of fragments} = \frac{Z_1 Z_2 e^2}{4\pi\epsilon_0 D},
\label{611}
\end{equation}

where $Z_1,Z_2$ are the atomic numbers of the two positively charged fragments (we shall temporarily assume that the kinetic energy of the neutrons emerging from the reaction is negligible), $e$ is the elementary charge and $D$ is the distance between the electrostatic centers of the two fragments at the moment of separation. The value of $D$ is known to be strongly dependent on the state of the deformation of the nucleus\cite{Wilkins} before the occurrence of fission. Under laboratory conditions, the deformation of the nucleus is minimal and hence the distance $D$ is at a minimum; the TKE therefore has the highest possible value. In a nuclear explosion, however, the conditions are substantially  different. Since the energy density (that is, the energy per unit volume of the material) is very high, the nucleus is expected to be very highly excited and hence the distance $D$ will be at a maximum. Accordingly, the TKE should be expected to have the lowest possible value. We shall now obtain a numerical value for the TKE in this specific case (i.e., when the TKE is minimum), and demonstrate that this numerical value is in agreement with the quantity $\Delta m v^2$.\\
\\
In general, determining the deformation and hence the distance $D$ for a highly deformed nucleus cannot be expected to give a precise answer, since such deformation is very circumstantial. We can, however, obtain a precise estimate of the minimum Q-value of the reaction by applying Eq.(\ref{611}) to the nucleus {\em before} the occurrence of fission. According to the recent models\cite{Williams},\cite{David}, the two positive fragments can be regarded, before the occurrence of fission, as two coupled oscillators. In such a model, the potential energy of the oscillators is constantly transformed into kinetic energy and vice versa, under the influence of the strong nuclear force. The total mechanical energy of the oscillators at any time will be therefore equal to the peak Coulomb potential, that is, the potential at the nadir of the displacement. Hence, the sum $(\mbox{P.Energy} + \mbox{K.Energy})$ at the moment of separation of the oscillators (that is, when fission occurs) must be equal to the peak Coulomb potential. But since the TKE of the fragments after  the separation is observed in practice to be numerically in agreement with only the potential energy at separation, then we must conclude -if we take the coupled oscillators model as valid- that the internal kinetic energy gets totally consumed in equalizing, or ``breaking'', the strong nuclear force. If we now assume that fission will occur when this internal kinetic energy is at a maximum, then calculating this quantity will be in fact quite simple. This quantity will be given by

\begin{eqnarray}
\lefteqn{\mbox{K.Energy (internal, max.)} = } \quad \nonumber\\
 & & \mbox{P.Energy (max.)} - \nonumber\\
 & & \,\,\, \mbox{P.Energy (at separation)} = \nonumber\\
 & & \frac{Z_1 Z_2 e^2}{4\pi\epsilon_0} 
\left( \frac{1}{D_{min}} \right. - \left. \frac{1}{D_{sep}} \right), 
\label{611b}
\end{eqnarray}

where $D_{min}$ is the minimum distance between the oscillators (or the distance at the nadir) and $D_{sep}$ is the distance at the moment of separation. The observed TKE of the fragments after separation typically indicate that fission occurs at a distance $D_{sep}$ that is larger than the nuclear diameter, or larger than 10fm (note: 1fm = $10^{-15}$ meters). The choice of $D_{min}$ for a highly deformed nucleus is actually a straightforward choice. Since the range of the strong nuclear force is known to be about 1.2fm (beyond that range, the nucleus will be unstable \cite{Williams},\cite{David}), the maximum acceptable value for $D_{min}$ will be therefore 1.2fm. Now, by comparing $D_{min}$ to $D_{sep}$, we can clearly see from Eq.(\ref{611b}) that the maximum internal kinetic energy will be approximately equal to the maximum potential energy. If we assume that the masses of the two oscillators -or coupled fragments- are $m_1$ and $m_2$, Eq.(\ref{611b}) can therefore be written as: K.Energy (internal, max.) $\approx$ P.Energy (max.), or,

\begin{equation}
\frac{1}{2} \: (m_1 + m_2) v^2 \approx  
\frac{Z_1 Z_2 e^2}{4\pi\epsilon_0 D_{min}} 
\label{612}
\end{equation}

The most common values for the atomic numbers $Z_1$ and $Z_2$ in the fission of $U^{235}$ are $Z_1,Z_2 = 54,38$. The distance $D_{min}$ is $\approx$1.2fm as we noted. By substituting with the numerical values of these variables into Eq.(\ref{612}) we obtain    

\begin{equation}
\frac{1}{2} \: (m_1 + m_2) v^2 \approx 3.93\times 10^{-10} \mbox{Joules.} 
\label{613}
\end{equation}

Now we note that the sum of the masses $(m_1+m_2)$ of the two main fragments will be, within numerical accuracy, approximately equal to the mass $m$ of the $U^{235}$ nucleus. This conclusion can be made by observing that the mass deficit in the fission of $U^{235}$, $\Delta m$,  is approximately 200Mev$/c^2$, or $3.55\times 10^{-28}$ Kg. Now the mass imbalance relationship in the reaction is as follows\footnote{Note that the average number of free neutrons released from the reaction is 2.5 neutrons, while there is one incident neutron. The average difference is therefore 1.5 neutrons.}:

\vspace*{-0.25in}

\begin{eqnarray}
\Delta m & \approx & m - (m_1 + m_2) - \nonumber\\
 & & \quad (\mbox{1.5 neutron rest-masses}), \nonumber\\
 & & 
\label{614}
\end{eqnarray}

\vspace*{-0.25in}

hence

\vspace*{-0.25in}

\begin{eqnarray}
(m_1 + m_2) & \approx & m - \Delta m - \nonumber\\
 & & \quad (\mbox{1.5 neutron rest-masses}) \nonumber\\
 & \approx & m - 3 \times 10^{-27} \mbox{Kg}. \nonumber\\
 & & 
\label{615}
\end{eqnarray}

Since the mass $m$ of the $U^{235}$ nucleus is $3.88 \times 10^{-25}$ Kg, we can clearly see that $(m_1 + m_2)$ can be taken to be approximately equal to $m$, within the numerical accuracy of the calculations. From Eq.(\ref{613}), we therefore find that the maximum velocity $v$ of the oscillating fragments is: $v= 45\times 10^6$ meters/sec. This value also represents the velocity that will be acquired by the vanishing mass $\Delta m$ at the moment of the separation (note that the velocity of the fragments after separation will be different from $v$, as we indicated above, since a new kinetic energy will be acquired from the energy released in the reaction). The minimum energy released in the fission of one atom of $U^{235}$ will be therefore given by

\begin{eqnarray}
Q_{min} = \Delta m v^2 & = & 3.55\times 10^{-28} \times (45\times 10^6)^2 \nonumber\\
                 & = & 7.2 \times 10^{-13} \; \mbox{Joules.} \nonumber\\
 & & 
\label{616}
\end{eqnarray}

Notice that this figure is approximately 4.5Mev, which is substantially less than the maximum Q-value of 200Mev predicted by the quantity $\Delta m c^2$. Of course, as $\Delta m v^2 \rightarrow \Delta m c^2$ (under ideal conditions), then $Q \rightarrow Q_{max}$, or 200Mev. To understand this result graphically, a typical distribution of the TKE of the fission fragments is shown in Fig.1 below (see refs.\cite{Hambsch},\cite{Thierens}). This graph represents data taken under laboratory conditions. Notice the Gaussian-like distribution of the TKE of the fragments, and notice that $Q_{max}$, or the quantity $\Delta m c^2$, occurs as a maximum upper limit only. This distribution, of course, supports the variable quantity $\Delta m v^2$, not the rigid quantity $\Delta m c^2$, which represents only one value (that is, only one data point) in the graph. While this graph shows a minimum Q-value of about 140Mev, this is due to the fact that data was collected under ideal conditions. If the atoms become very highly excited (which occurs in a nuclear explosion), this plot is expected to shift toward lesser values of the TKE, as we indicated earlier (and in fact may not become Gaussian in shape). If the distance $D_{min}$ in Eq.(\ref{612}) is then stretched up to 1.2fm for all the atoms, then this distribution will be in fact compressed to one vertical line, representing a minimum Q-value of 4.5Mev as we calculated.  

\par
\vspace*{.15in}

\psfig{figure=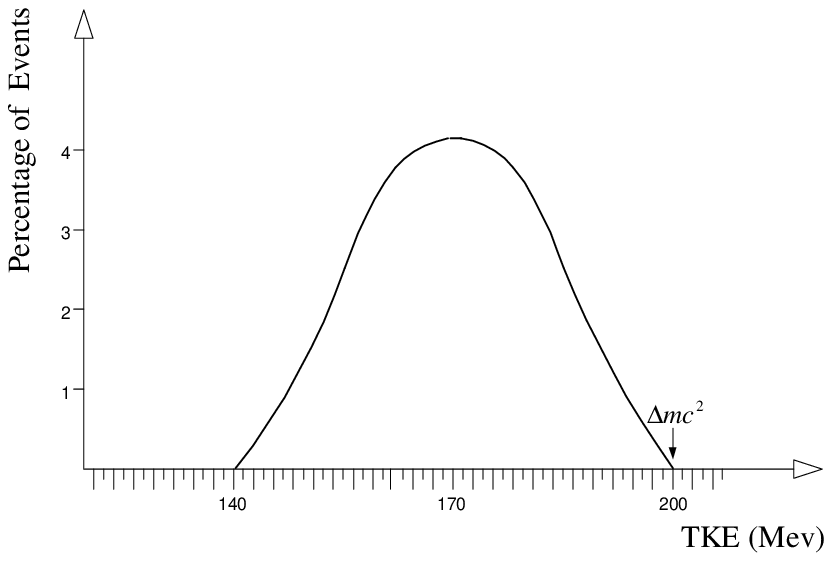}

\vspace*{.1in}

{\small Figure 1: Typical plot of the distribution of the total kinetic energy of the fission fragments of either $U^{235}$ or $Pu^{239}$ under laboratory conditions. The Gaussian-like distribution of the energy supports the variable quantity $\Delta m v^2$, not the rigid quantity $\Delta m c^2$, which occurs as a special case and an upper limit only.}\\
\\
{\bf 6.2. Application of the minimum Q-value to the Little Boy example:}\\
\\
We shall now proceed to apply the minimum Q-value that we calculated, $Q_{min}$, to the case of the Little Boy bomb; but we shall first demonstrate that we can obtain approximately the same numerical value by using a totally different approach. Serber\cite{Serber}  had given a phenomenological formula for calculating the mean velocity of the fragments after separation, $v^{\prime}$, that was derived by using the kinetic theory of gases. That formula is specific to nuclear explosion models, and gives the velocity as follows:

\pagebreak

\begin{equation}
v^{\prime}  = \frac{1}{2} \: \delta \: R_c \: \frac{N}{\tau},
\label{617}
\end{equation}

where:

\begin{itemize}
\item $N$ is the effective neutron number (or neutron multiplication factor) and is approximately equal to 1;
\item $\tau$ is the mean time between fissions and is equal to $10^{-8}$ seconds;
\item $R_c$ is the critical radius and is equal to 9cm or $9 \times 10^{-2}$ meters;
\item $\delta$ is a quantity given by $\delta = (M/M_c)^{\frac{1}{3}} - 1$, where $M/M_c$ is the ratio of the total mass of $U^{235}$ used to the critical mass. This ratio was 3 in the Little Boy, hence $\delta = 0.44$.
\end{itemize}

Using the above numbers in Eq.(\ref{617}) gives $v^{\prime}  = 1.98 \times 10^6$ meters/sec. If we now apply Eq.(\ref{24}) to the fission fragments, noting that the mass-energy quantity is approximately equal to $\frac{1}{2} \: m_0 v^2$ at sub-relativistic velocities, then we have

\begin{eqnarray}
\lefteqn{\mbox{TKE}\; (\mbox{fragments}) = }\qquad \nonumber\\
 & & \frac{1}{2} (\sum m_{\mbox{\small frag}}) 
v^{\prime 2}  \approx   \frac{1}{2} \: m v^{\prime 2} \nonumber\\ 
 & & \approx  \frac{1}{2} \times 3.88 \times 10^{-25} 
\times (1.98 \times 10^6)^2 \nonumber\\
 & & = 7.6 \times 10^{-13} \; \mbox{Joules/atom.}
\label{618}
\end{eqnarray}

This estimate, though not accurate, is of course very close to the minimum Q-value given by Eq.(\ref{616}). We shall now use the minimum Q-value to obtain an estimate for the total amount of energy released in the explosion of the Little Boy. To calculate that quantity, we note that there are $2.58 \times 10^{24}$ atoms per Kg of $U^{235}$ and that a total of 40Kgs were used in the bomb; hence,

\begin{eqnarray}
\lefteqn{\mbox{E(total)}  = }  \nonumber\\
 & & \!\!\!\!\!\!\! (7.2 \times 10^{-13} \mbox{J/atom}) \times (2.58 \times 10^{24}) \times 40   \nonumber\\
 & & \!\!\!\!\!\!\! = 7.43 \times 10^{13} \mbox{J.} \nonumber\\
 & &
\label{619}
\end{eqnarray}

Given that 1 ton of TNT is equivalent to $3.6 \times 10^9$ Joules, we can calculate the maximum expected yield of the bomb in tons:

\begin{eqnarray}
\mbox{E(tons)} = \frac{7.43 \times 10^{13}}{3.6 \times 10^9} & = & 20,640 \: \mbox{tons} \nonumber\\
 & \approx & 20.6 \: \mbox{Ktons}. \nonumber\\
 & & 
\label{620}
\end{eqnarray}

This number is of course very close to the actual reported yield of 15Ktons for the Little Boy\cite{Serber} (we must point out here that the $U^{235}$ used in that application was only 90\% enriched, which makes the effective mass about 36Kgs. This actually brings the two numbers much closer to each other).\\
\\
To compare the above result with the original calculation, we first note from Fig.1 that the average energy release per atom was estimated to be about 170Mev under laboratory conditions. Had we used that value, we would have obtained a theoretical yield of 780Ktons! To compare the efficiency in both cases, we note that according to this calculation the lowest estimate for the efficiency of the bomb will be

\begin{equation}
\mbox{Efficiency} = \frac{15\mbox{Ktons}}{20.6\mbox{Ktons}} \approx 73\%,
\label{621}
\end{equation}

\vspace*{0.125in}

which is in sharp contrast with the figure of $2\%$ that resulted from the original calculation (that is, $15/780 = 2\%$). The important conclusion here of course is not the efficiency figure, but the fact that the theoretical yield of 20.6Ktons is in very good agreement with the actual yield of 15Ktons, as opposed to the original expectation of 780Ktons.\\
\\
It will be therefore apparent from the above analysis that the Q-value of a nuclear reaction, as given by the quantity $\Delta m v^2$, is in very good agreement with the experimental data. The Q-value given by the quantity $\Delta m c^2$ is therefore a maximum theoretical limit. Moreover, it will be clear that the Q-value of any given reaction is not a constant, but is a function of the state of deformation of the nucleus.\\
\\
\underline{Remark about reference\cite{Serber}:}\\
It is interesting to note that Serber indicated that he had substantial arguments with Oppenheimer and his other colleagues about the assumptions that were used in calculating the efficiency of the bomb. Moreover, it is evident from the book that he contradicted his own opinion a number of times, stating in some instances that: ``it is just possible for the reaction to occur to an interesting extent before it is stopped by the spreading of the active material". This remark was made in light of the calculation that showed that the entire reaction should take less than $10^{-6}$ sec. for completion. Of course, from the calculation of the velocity of the fragments given above, Serber and his colleagues were aware that the expansion in the radius of the core should be less than 2 meters during the $1\mu s$ that is required for completion of the reaction. Further, it is interesting to note that other mechanisms of fission weren't factored in at all in the calculation (such as gamma-induced fission;  spontaneous fission as a result of the state of very high excitation of the atoms, etc).\\
\\
{\large\bf 7. A note on the nature of beta-decay and the problem of the solar neutrinos:}\\
\\
We shall now explain briefly how the equation $H=mv^2$ can provide a satisfactory explanation for the experimental data observed in the beta-decay process, without the necessity of having a second particle, namely, the neutrino, emitted in the process. This by no means is meant to suggest that the neutrino does not exist, since the existence of the particle was verified experimentally. What this discussion will demonstrate, however, is that neutrinos are likely to be generated in high-energy nuclear reactions, but not in the radioactive beta-decay of atoms. A possible explanation for the phenomenon of the shortfall in the solar neutrinos will also be given.\\
\\
{\bf 7.1. The beta-decay spectrum:}\\
\\
In Fermi's original theory\cite{Perkins1}, the principle of conservation of energy, linear momentum, and angular momentum require the existence of a second particle, that is, the neutrino. By using this two-particle model, Fermi concluded that the number of electrons $N$ emitted for each momentum value $P_e$ will be given by

\begin{equation}
N \propto P_e^2 \: (E_{max} - E_k)^2,
\label{63}
\end{equation}

where $E_k$ is the electron's kinetic energy and $E_{max}$ is the maximum possible energy (as calculated from the traditional expression $H=\Delta mc^2$). The Kurie plot that is usually obtained in the beta-decay experiments confirms Eq.(\ref{63}). Fermi's theory does not exclude, but in fact it does require, that the linear and angular momenta of the decaying nucleus (or neutron) be factored in the momentum conservation equations. If we now assume that a neutrino will not be emitted, Fermi's theory and Eq.(\ref{63}) still hold since the second particle in Fermi's model will be the decaying nucleus itself (that is, the two-particle model is still applicable). The problem in this case, however, becomes the principle of conservation of energy, since basic mechanics imply that nearly all the missing energy must be carried away by the electron's kinetic energy. Let us consider the specific case of the beta-decay of a nucleus (such as the tritium nucleus, which releases a maximum energy of 18.6Kev). According to the theory of Quantum Chromodynamics (QCD), the process of beta-decay occurs according to the following reaction:

\begin{equation}
d \rightarrow u + e^{-} +  \bar{\nu_e},
\label{64}
\end{equation}

where $d$ is a down quark, $u$ is an up quark, $e^{-}$ is the emitted electron, and $\bar{\nu_e}$ is an antineutrino. If we now assume that no neutrino emission will occur, then the kinetic energy of the emitted electron, $E_k$ (which must be equal to the Q-value of the reaction\cite{Perkins1},\cite{Williams}), will be given by

\vspace*{-0.25in}

\begin{eqnarray} 
E_k & = &  H(d) - H(u) \nonumber\\
    & = &  m_d \: v_d^2 - m_u \: v_u^2,
\label{65}
\end{eqnarray}

where $m_d$ and $m_u$ are the relativistic masses of the down and up quarks, respectively, and $v_d,v_u$ are their respective velocities inside the nucleus. Since it is practically observed that the recoil energy of the nucleus is negligible, then the kinetic energies of the down and up quarks (that is, the kinetic energies before and after the emission) must be approximately equal. Further, since the rest masses of the down and up quarks are known to be very nearly equal\cite{Williams}, then the velocities $v_d$ and $v_u$ can be considered to be approximately equal without introducing a significant error. In addition, the difference in the relativistic masses $(m_d - m_u)$ must be equal to the apparent mass deficit $\Delta m$ of the nucleus. Hence Eq.(\ref{65}) can be reduced to 

\vspace*{-0.25in}

\begin{eqnarray}
E_k & \approx & (m_d - m_u) \: v^2 \nonumber\\
    & \approx & \Delta m v^2,
\label{66}
\end{eqnarray}

where $v$ is the velocity of either quark inside the nucleus. We can now clearly see that the electron's kinetic energy will be a variable that depends on the quark's velocity. If $v \rightarrow c$, then $E_k \rightarrow \Delta m c^2$. Usually, quark momenta are observed experimentally by means of transverse hadron jets in detectors like the CERN UA2 detector. Observance of such jets did in fact confirm that quark velocities will vary from $0$ up to values very close to $c$. Moreover, plots of transverse hadron momenta\cite{Hanson},\cite{Tasso} show very strong resemblance with the kinetic energy plots usually obtained in the beta-decay experiments. These facts of course lead to the conclusion that Eq.(\ref{66}) is in agreement with the  experimental beta-decay data; that is, the fact that $E_k$ does vary from $0$ up to the theoretical maximum $\Delta m c^2$ with a distribution that does correlate with the distribution of quark velocities (a typical plot is shown in Fig.2 below) is a fact that can be seen from Eq.(\ref{66}) and from the above discussion. These facts also explain why Eq.(\ref{63}) is valid, even though we have assumed that no neutrino is present. 

\par
\vspace*{.15in}

\psfig{figure=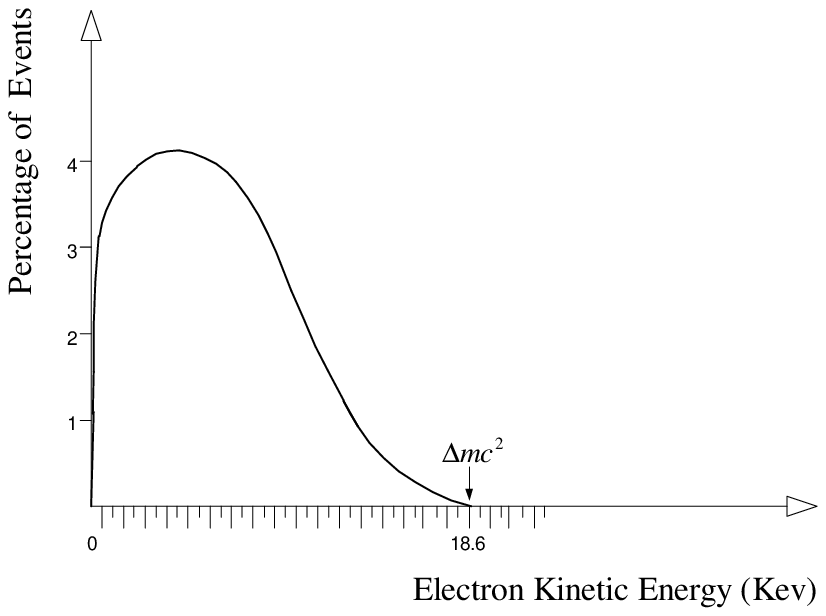}

\vspace*{.1in}

{\small Figure 2: Typical plot of the distribution of the electron's kinetic energy in beta decay. As in the case of nuclear fission, the quantity $\Delta m c^2$ appears as a special case and an upper limit only. The distribution can be understood in terms of the quantity $\Delta m v^2$, rather than the hypothesis of $\Delta m c^2$ minus a neutrino.}\\
\\
The above result shows that a neutrino is not likely to be generated in the ordinary process of radioactive beta-decay. The situation is different, however, in high-energy nuclear reactions. In the sun, for example, neutrinos are thought to be generated in the principal reaction\cite{Bahcall}

\begin{equation}
p+p \rightarrow \mbox{ }^2H + e^{+} + \nu_e
\label{67}
\end{equation}

We must now realize that such a reaction cannot be understood by simply comparing the rest-masses of the particles involved. A more accurate representation of the reaction, that takes velocities into account, will be written as

\begin{equation}
2m_p v_p^2 = m_H v_H^2 + m_e v_e^2 + \mbox{Missing Energy},
\label{68}
\end{equation}

where the subscripts $p, H$ and $e$ denote the proton, the deuteron and the positron, respectively. Unfortunately, unless the velocities of those particles are determined with a reasonable degree of accuracy, calculating the amount of missing energy will be very difficult. If in fact the equation does balance without the missing energy term, then no neutrino will result. Otherwise, the missing energy will appear in the form of a neutrino.  This can be a very plausible explanation for the phenomenon of the shortfall in the solar neutrinos. Moreover, if a neutrino does in fact emerge in every reaction, then the energies of those neutrinos can have a very wide distribution (as some very recent experiments seem to suggest). Of course, more work needs to be done in the future to obtain accurate estimates of the velocities of the particles involved in the solar reactions.\\
\\
{\bf 7.2. The problem of angular momentum conservation:}\\
\\
The above discussion resolves the apparent mystery of the lost energy in the process of radioactive beta-decay. One question that remains unanswered, however, is the apparent violation of the principle of conservation of angular momentum. Since we have excluded the neutrino as an unnecessary proposition in beta-decay, then conservation of total angular momentum will be violated since, for most reactions, the angular momentum of the decaying particle is different from the sum of the angular momenta of the emerging particles (for example, the angular momentum of the tritium nucleus is $\frac{1}{2}\hbar$, that of the helium nucleus is $\frac{1}{2}\hbar$, and that of the electron is also  $\frac{1}{2}\hbar$).\\
\\
To understand this apparent violation, we must understand what is meant by the ``principle of conservation of angular momentum''. From elementary mechanics, the total angular momentum $P_\phi$ of a system of particles is conserved if the net external torque $T_{ext.}$ acting on that system is zero; that is, if

\begin{equation}
T_{ext.} = \frac{dP_\phi}{dt} = 0,
\label{69}
\end{equation}

which implies that $P_\phi$ must be a constant. Since the introduction of the neutrino hypothesis by Pauli, the prevailing assumption was that no external torques (or angular forces) are present to act on the particles involved in the beta-decay process, and hence the total angular momentum must be conserved. The conclusions that emerged from the theory of QED later, however, make that assumption very doubtful. A decaying nucleus is not an isolated system. It is acted upon by the orbiting electron(s) and by its own field. The influence of such factors on the angular momentum of the nucleus and/or the emerging electron have not been studied. While such a study will be beyond the objective and scope of this paper, the point that must be emphasized here is that the angular momentum of a nucleus must not necessarily be conserved in the process of beta-decay.\\
\\
\\
{\large\bf 8. Summary and conclusions:}\\
\\
From the presented analysis, there is little evidence to argue against the fact that wave mechanics predicts that the total energy of a particle must be given by the quantity $m v^2$. On the other hand, we have clearly seen that Einstein's relativistic treatment of energy can be interpreted in two ways, depending essentially on how the integration of the basic kinetic energy equation is performed. According to the procedure that Einstein used in the integration, he was led to the formula $E_k = mc^2 - m_0 c^2$, from which he concluded that the total energy is $mc^2$. Doing the integration differently, however, leads us to the conclusion that the total energy is $m v^2$, while maintaining the numerical value of Einstein's integral. These facts therefore suggest that, for wave mechanics to be consistent with relativity, the expression $H=mv^2$ must be used as the total energy formula. We have shown that this formula can correct a substantial problem in Dirac's theory of the electron, specifically, the fact that Dirac's calculation of the velocity resulted in a value that does not correlate with physical reality. We have also shown that this formula results in a theoretical prediction of the energy released from nuclear fission that is in very good agreement with the actual measured value. This last fact is a proof that the formula is indeed distinguishable from currently known experimental data. The other problems that were enumerated in Sec.1 can also be now understood in light of this result. We have commented on the process of beta-decay and offered a very plausible explanation for the phenomenon of the shortfall in the solar neutrinos.\\
\\
Having reached the above conclusion, we must not forget, however, that there is a sizeable record of experimental evidence that does support the predictions of the theory of special relativity \cite{Bucherer},\cite{Ives},\cite{Rossi},\cite{Hafele}. The author fully supports the theory, as far as those experimental facts are considered. For instance, the principles of relativity of time, relativity of mass, etc., are all valid principles in the opinion of the author. What is being questioned here is the old understanding of the mass-energy equivalence principle in view of the reasons mentioned above.\\
\\
\\
{\large\bf Acknowledgements:}\\
\\
The author is grateful to Prof. Howard Haber of the University of California, Santa Cruz, for his valuable advice in submitting this manuscript. The author also thanks Dr. Walter Becker, retired, and an anonymous referee for reading the paper in detail and suggesting improvements.

\end{document}